\documentclass[prl,twocolumn,showpacs,superscriptaddress,nofootinbib,preprintnumbers]{revtex4}

\usepackage{amsfonts}
\usepackage{graphicx}
\usepackage{amsmath}
\usepackage{bm}

\newcommand*{\chpt}{\raise0.4ex\hbox{$\chi$}PT}

\newcommand{\Delstar}{\ensuremath{\Delta^{\raise0.18ex\hbox{${\scriptstyle *}$}}}}
\def\gtwid{{\,\raise.35ex\hbox{$>$\kern-.75em\lower1ex\hbox{$\sim$}}\,}}
\def\ltwid{{\,\raise.35ex\hbox{$<$\kern-.75em\lower1ex\hbox{$\sim$}}\,}}
\def\leftvec{{\raise1.5ex\hbox{$\leftarrow$}\kern-1.00em}}
\def\rightvec{{\raise1.5ex\hbox{$\rightarrow$}\kern-1.00em}}
\def\half{{\scriptstyle \raise.2ex\hbox{${1\over2}$}}}
\def\threehalves{{\scriptstyle \raise.15ex\hbox{${3\over2}$}}}
\def\third{{\scriptstyle \raise.15ex\hbox{${1\over3}$}}}
\def\third{{\scriptstyle \raise.15ex\hbox{${1\over3}$}}}
\def\twothirds{{\scriptstyle \raise.15ex\hbox{${2\over3}$}}}
\def\fourth{{\scriptstyle \raise.15ex\hbox{${1\over4}$}}}

\newcommand*{\bea}{\begin{eqnarray}}
\newcommand*{\eea}{\end{eqnarray}}
\newcommand*{\be}{\begin{equation}}
\newcommand*{\ee}{\end{equation}}

\begin{document}

\preprint{JLAB-THY-08-910}

\title{Magnetic Moments of $\Delta$ and $\Omega^-$  Baryons with Dynamical Clover Fermions}

\author{C.\ Aubin}
\author{K.\ Orginos}
\affiliation{Department of Physics, College of William \& Mary, Williamsburg, VA 23187}

\author{V.\ Pascalutsa}
\author{M.\ Vanderhaeghen}
\affiliation{Institut f\"ur Kernphysik, Johannes Gutenberg-Universit\"at, D-55099 Mainz, Germany}

\begin{abstract}
We calculate the magnetic dipole moment of the $\Delta$(1232) and $\Omega^-$
baryons with 2+1-flavors of clover fermions on anisotropic lattices using a background magnetic field. 
This is the first dynamical calculation of these magnetic moments using a background field technique.
The calculation for $\Omega^-$ is done at the physical strange quark mass, with the result in units of the physical nuclear magneton $\mu_{\Omega^-}= -1.93(8)(12)$ (where the first error is statistical and the second is systematic)
compared to the experimental number: -2.02(5). 
The $\Delta$ has been studied at three unphysical quark masses, corresponding to pion mass $m_\pi = 366, 438, $ and $548$ MeV. The pion mass dependence is compared
with the behavior obtained from chiral effective field theory.
\end{abstract}


\maketitle

Calculations of hadron properties from first principles using lattice QCD have been rapidly
advancing in recent years.   
The newly available fully-dynamical (unquenched) lattice configurations 
have made it possible to significantly reduce the systematic error of lattice calculations.
Of the properties that can now be reliably computed on the lattice are the electromagnetic 
(e.m.) properties
of baryons and in particular their electromagnetic moments.  
Here we present a first dynamical calculation of the magnetic dipole moment of the $\Delta(1232)$ and $\Omega^-(1672)$ baryons using a background e.m.\ field. 

These particular baryons are chosen for the following reasons. They both are distinguished
members of the baryon decuplet and as such they have much in common. On the other hand,
while the magnetic moment of the
$\Omega$ is measured to a few-percent accuracy, the tiny lifetime of the $\Delta$ resonance
($\approx 6\times 10^{-24}$ sec) hinders the determination of its magnetic
moment and the experimental efforts are still ongoing. This is why a simultaneous 
lattice calculation for these two baryons can both be tested against experiment 
in the $\Omega$ case and provide predictions in the case of the $\Delta$.

The background-field method adopted here 
is presently the simplest and cleanest way to access the static
e.m.\ moments on the lattice, as  it amounts  simply to measuring the shift in the mass 
spectrum upon applying a classical background field~\cite{Bernard:1982yu} 
(for most recent applications to baryons in the quenched approximation see Refs.~\cite{Cloet:2003jm,Lee:2005ds}).  
 The other possibility is the form factor calculation extrapolated to the $q^2=0$ point (from the minimum momentum-transfer on the lattice, which is $2\pi/aL$, with $L$ the number of points in the spatial direction). However, in comparison with the background-field method, 
this method is additionally complicated by the noise of the three-point function calculation as well as the uncertainties in the $q^2$ extrapolation, see Refs.~\cite{Alexandrou:2007we,Alexandrou:2008bn} for recent calculations of the $\Delta$ e.m.\  form factors (the first
calculation is done in the quenched case, and the latter in the dynamical case).

In order to calculate the magnetic dipole moments, we implement
the constant background magnetic field in the following fashion. 
 On a given configuration, we multiply all of the $SU(3)$ gauge fields by a $U(1)$ gauge field, and invert the Dirac operator on that background to get the quark propagator in a background field. The $U(1)$ links are given by
\be
	U_\mu(x) = \exp\left[iq a A_\mu(x)\right] \ ,
\ee
where $q$ is the charge of the quark whose propagator we are calculating. For a constant magnetic field with a magnitude of $B$ pointing in the $+z$-direction, the usual choice is
$A_\mu(x,y,z,t)= a B x\,\delta_{\mu y}$. The problem with this choice is that due to the condition that the gauge links $U_\mu$ must be periodic, the field is continuous only if $q a^2 B=2\pi n/L$, with integer $n$. Hence,
the minimal value of $B$ is severely limited by the size of the lattice.
This limitation is somewhat relaxed for the following choice of the 
field~\cite{Damgaard:1988hh,Rubinstein:1995hc,Aubin:2008hz}:
\bea\label{eq:y-link-mod}
	A_\mu(x,y,z,t) & = &
	\begin{cases}
	a B x\; \delta_{\mu y} & \text{if } x\ne L-1\\
	-a B L y\; \delta_{\mu x} & \text{if } x= L-1\ .
	\end{cases}
\eea
Thus, all of the $y$-links are modified by $\exp\left[i q a^2 B x\right]$, all $x$-links on the $x$ boundary are modified by $\exp\left[-i q a^2 B Ly \right]$, and all other links are unchanged. The additional modification of the links on the $x$-boundary, allows us to achieve continuous 
constant field everywhere on the lattice with a more relaxed constraint
on the value of the field:
\be\label{eq:quant}
	qa^2 B = \frac{2\pi n}{L^2}.
\ee
The latter periodicity constraint corresponds with the more physical requirement that the magnetic flux
(plaquette) remain continuous through the boundary. 

With this \cite{Bernard:1982yu}, one can calculate a baryon two-point function which behaves for large time in the usual manner
\be\label{eq:2pt}
	C(t) \sim A\ e^{-m(B)\, t} + \ldots\ ,
\ee
but with the exponential damping governed by a $B$-field dependent mass
\be\label{eq:mB}
	m(B) = m_B - \mu_z B + O(B^2)\ .
\ee
Here, $m_B$ is the mass of the baryon in the absence of any external field. The magnetic moment 
along the direction the field is given by
\be\label{eq:magmom}
	\mu_z = \mu \,S_z/S \ ,
\ee
where $\mu$ is the value of the magnetic moment, $S_z$ is the spin projection and 
$S$ the total spin (in our case $S=3/2$ and so $S_z$ can take the values $\pm 3/2$ and $\pm 1/2$).
It is useful to form the quantity 
\be
\Delta m_{S_z} = m(-B)-m(B) = 2 \mu B\mbox{$\frac{S_z}{S}$} + O(B^3)
\ee
to cancel the effect of the next order in the $B$ expansion. We have computed this quantity for all the various
spin-projection values and extracted the magnetic moment from the following combination:
\be\label{eq:magmoment}
	\mu = 
	\frac{1}{8 B} \biggl[
	 \Delta m_{3/2}
	- \Delta m_{-3/2}+3(\Delta m_{1/2}- \Delta m_{-1/2}) \biggr].
\ee
Since the $\Delta m_{S_z}$ are highly correlated, the error on $\mu$ is determined using a jackknife, and this combination is chosen to average the results from all spin components under the jackknife procedure. 

On a technical note, the number input into the simulation is $qa^2B$, and thus includes the product of the quark charge and the magnetic field in lattice units. In order to account for the quark charges of the up, down, and strange quarks, for a single magnetic field $B$ we must use two values of $q a^2 B$, corresponding to the fact that $q_u = -2q_{d,s}$. For particles made up with only a single quark ($\Delta^{++},\Omega^-$), we need one input value, but for the $\Delta^+$ or the nucleon, for example, we must use two inputs that differ by a factor of $-2$ so that the quarks all experience the same $B$-field. 

Notice that even with the modified periodicity constraint in Eq.~(\ref{eq:quant}), the minimum value of the magnetic field may still be large enough to distort the baryons, and thus introduce errors into the extrapolation of the magnetic moment. Ideally we would use volumes large enough that this would not be true, however this can become rather expensive. Earlier studies \cite{Bernard:1982yu,Lee:2005ds}, ignore the periodicity constraint,\footnote{These studies do not include the modification of the $x$-links on the boundary, and thus their periodicity constraint is given by $qa^2 B = 2\pi n / L$, an order of magnitude larger than ours.} using small fields that would not distort the particles and also ensure the linear relationship between the extracted mass in Eq.~(\ref{eq:mB}) and the magnetic field. In addition, they imposed Dirichlet boundary conditions and place the source in the center of the lattice to perhaps ensure the quarks will not feel the effects of the discontinuity.

The difficulty with this approach, however, is that there are significant finite volume effects in the results in the magnetic moments, using their implementation. Specifically, in the quenched calculation of Ref.~\cite{Lee:2005ds}, the authors see effects that are as large as 35\% for the lightest pion mass when comparing the $16^3$ volume to a $24^3$ volume, with a lattice cutoff of $a^{-1}\approx 2\ {\rm GeV}$, and a pion mass of about 522 MeV. Since taking the pion mass closer to the physical point, finite volume errors become more substantial, we would like to reduce the finite volume effects coming from the background field as much as possible.

Using the implementation of the magnetic field above, we have shown that ignoring the periodicity constraint somewhat will not introduce noticeable finite volume errors coming from the background field, so long as one uses the implementation of the background field shown in Eq.~(\ref{eq:y-link-mod}) \cite{Aubin:2008hz}. With this method, we are able to trust results coming from simulations on smaller volumes
(still keeping $m_\pi L\gtwid 4$ so we can minimize finite volume effects coming from a small pion mass), which are less expensive. Using the methods in Ref.~\cite{Lee:2005ds}, for example, one is restricted only to larger volumes.

We now present our results, which are the first dynamical 
calculations for magnetic moments using a background field.

We use dynamical anisotropic lattices with 2+1 flavors of Stout-smeared Clover fermions \cite{Edwards:2008ja,Lin:2008pr}, on two volumes and a single lattice spacing. We show the relevant parameters in Table~\ref{tab:params}. Note that on these lattices, a bare quark mass parameter of $-0.0743$ corresponds to the physical strange quark mass. Both sets of configurations have an inverse spatial lattice spacing of 1.61 GeV and an anisotropy of about 3.5 (so $a_t^{-1}\approx 5.61$ GeV). More complete information on these configurations, specifically the tuning of the lattice parameters, can be found in \cite{Edwards:2008ja,Lin:2008pr}.

\begin{table}[t]
\caption{Lattice parameters used for the current work. On these lattices, the anisotropy is $a_t/a_s\approx 3.5$, and we show the pion masses on the lattices in physical units. The $a_s m_{\rm val} = -0.0743$ dataset was used to calculate the $\Omega^-$ magnetic moment, so the $m_{\Delta,\Omega}$ listed in that row is the $\Omega^-$ mass, while the other baryon masses are those of the $\Delta$.}
\begin{center}
\begin{tabular}{ccccc}
\hline
Volume & $a_t m_{\rm val}$ & $m_\pi$
& $m_{\Delta,\Omega}$ & \# \\
& (GeV)  & (MeV) 
& (GeV) & configs\\\hline
$16^3\times 128$ & -0.0808 & 548 & 1.562 & 110\\
$16^3\times 128$ & -0.0830 & 438 & 1.485 & 91\\\hline
$24^3\times 128$ & -0.0840 & 366 & 1.408 & 202\\
$24^3\times 128$ & -0.0743 & 366 & 1.65 & 213\\
\hline
\end{tabular}
\end{center}
\label{tab:params}
\end{table}%

In all $\Delta$ cases we used three magnetic fields for the simulations, corresponding to $n=\pm1/2, \pm1, $ and $\pm2$ in Eq.~(\ref{eq:quant}). For the $\Omega$, we only used $n=\pm1/2$ and $n=\pm1$. The $n=\pm1/2$ field does not satisfy the periodicity constraint. We expect the errors entering here due to this to be negligible as was shown in Ref.~\cite{Aubin:2008hz}. Even with these fields, we already see higher-order terms appearing in the expression for the masses extracted from two-point functions, so larger magnetic fields will begin to introduce effects coming from even higher-order terms in Eq.~(\ref{eq:magmoment}). We calculate all four spin projections for the baryons, as well as using both positive and negative magnetic fields, and we average over all of these to reduce the errors. Additionally, on each configuration, we calculated the quark propagators starting from four time sources $t=0,32,64,$ and 96, using the EigCG algorithm developed in Ref.~\cite{Stathopoulos:2007zi}  to decrease significantly the time it takes to invert the Dirac operator.

\begin{figure}[t]
\begin{center}
\includegraphics[width=3.2in]{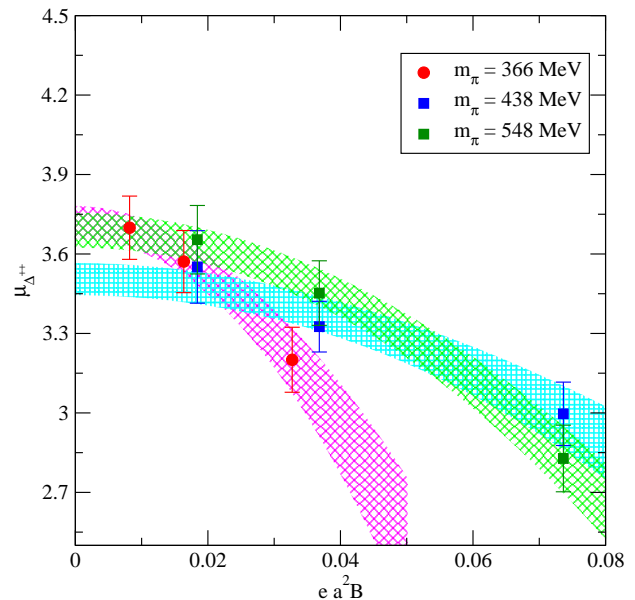}
\caption{In this figure we show the magnetic moment in units of the physical nuclear magneton for the three input magnetic fields used, as well as quadratic fits to each dataset, to remove residual $B^2$ dependence in the magnetic moments.}
\label{fig:deltamm}
\end{center}
\end{figure}

In Fig.~\ref{fig:deltamm} we show the $\Delta^{++}$ magnetic moments in units of the physical nuclear magneton $\mu_N$, for the three pion masses simulated. One can see noticeable effects coming in at $O(B^2)$, and we have fit each dataset to a quadratic form
\begin{equation}\label{eq:B2extrap}
	\mu = \mu_0 + b (ea^2 B)^2\ .
\end{equation}
With each set, since we only have two parameters and three data points, a correlated fit is not possible, and we do not take the fitted value of $\mu_0$ (in principle this would be the most appropriate value, as it subtracts out the $B^2$ dependence) and its error as our final result. Instead, we see that for the smallest of the $B$-fields simulated, the $O(B^2)$ effects are small, and so we take that data point as our determination of the magnetic moment (in fact, the data point at the smallest value of $B$ is consistent, as one can see from the Figure, with the value of $\mu_0$). The fits performed give drastically smaller errors than the data, and so we choose to use the error from the data, to account for possible uncertainties in this method. There is a slight shift (within errors) in the extracted $\mu_0$ compared with the smallest $B$-field data point, and for the heavier mass this is its largest at 5\%. Similar results can be seen for the $\Delta^+$ and $\Omega^-$, and we show all of our results in Table.~\ref{tab:results}. Note that our results for the $\Delta^-$ are not included, because with this method of calculation, we have the exact equality
$ \mu_{\Delta^{++}} = - 2 \mu_{\Delta^{-}}$.

\begin{figure}[t]
\begin{center}
\includegraphics[width=3.0in]{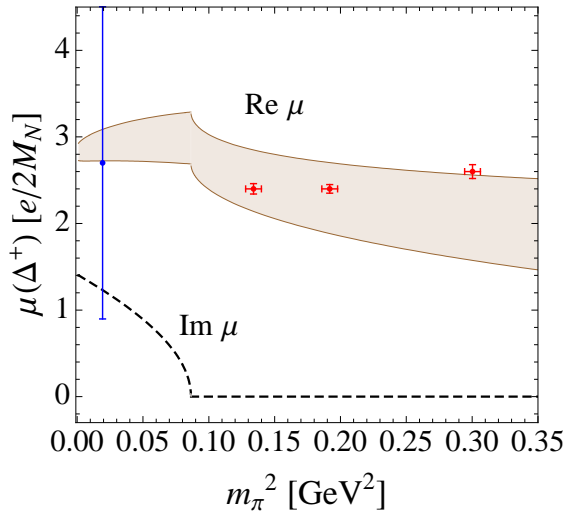}
\caption{ Chiral effective field theory 
calculations of Ref.~\cite{Pascalutsa:2004je} for the $m_\pi$ dependence 
of $\mu_{\Delta^+}$, in units of the physical nuclear magneton. 
Both real and imaginary parts of $\mu_\Delta$ are displayed. For the former, 
the bands show a theoretical error estimate, as described in the text. 
The value at the physical pion mass corresponds with the experiment 
of Ref.~\cite{Kotulla:2002cg}, where both statistical and
systematic errors are displayed.  
}
\label{fig:chpt}
\end{center}
\end{figure}

Also in the table, we show the experimental numbers for these quantities. We
can see that for the $\Delta^{++}$, there is a possible upward trend as we
decrease the pion mass. We illustrate this in Fig.~\ref{fig:chpt}, where we 
show the present results together with the chiral effective field theory 
calculations of Ref.~\cite{Pascalutsa:2004je} for the $m_\pi$ dependence 
of $\mu_{\Delta^+}$. These calculations have one free parameter for 
$\mu_{\Delta^+}$, corresponding with its value in the chiral limit.   
We also indicate a theoretical error band, corresponding with an 
error of $(m_\pi + m_\pi^{phys}) / m_\Delta^{phys}$ estimating 
the corrections of next chiral order, 
with $m_\pi^{phys}$ the physical pion mass 
and $m_\Delta^{phys}$ the physical $\Delta$ mass.   
One notices a strong cusp behavior for the real part of $\mu_\Delta$, which is 
due to the opening of the $\Delta \to \pi N$ decay channel. Therefore, 
no strong conclusions on the value of $\mu_\Delta$ at the physical point 
can be made until this extrapolation has been done. We leave such 
a systematic study for a future work.   
On the experimental side, there are new experiments  from MAMI for the 
magnetic moment of the $\Delta^+$ with much reduced errors, 
yet these have not yet been fully analyzed.

\begin{table}[t]
\caption{Calculated magnetic moments in units of $\mu_N$, the physical nuclear magneton (taken as the value for the data for the smallest $B$-field, as discussed in the text). For comparison, we have combined all experimental errors in quadrature.}
\begin{center}
\begin{tabular}{ccccc}
$m_\pi$ & $\mu_{\Delta^{++}}$  
& $\mu_{\Delta^{+}}$
&$\mu_{\Delta^{0}}$ & $\mu_{\Omega^-}$ 
\\\hline
548 & 3.65(13) 
& 2.60(8) & -0.07(2) & \\
438 & 3.55(14) 
& 2.40(5) & 0.02(3)&  \\
366 & 3.70(12) 
& 2.40(6) &0.001(16) &$-1.93(8)$\\
\hline
PDG: & 5.6(1.9)
& 2.7(3.5) & --- & $-2.02(5)$\\\hline
\end{tabular}
\end{center}
\label{tab:results}
\end{table}%

Since the sea quarks do not carry electric charge (which is the case for \emph{all} current lattice simulations), there is a relationship that holds within the quark model in the isospin limit, where
$\mu_{\Delta^{++}} = 2 \mu_{\Delta^{+}}$. 
Clearly this relationship does not hold with our results above, but we could use this relationship to reduce the systematic uncertainties in our determination. This would clearly increase the values obtained for the $\Delta^{++}$ and reduce it for the $\Delta^+$. 

As for the $\Omega^-$, the strange quark mass is close to its physical value (as we can see by the fact that the $\Omega^-$ mass is close to the observed value), so we expect the result to match more closely to the experimental value. As we can see, it agrees tremendously well. This agreement is expected, as quantities involving the $\Omega^-$ should have little dependence on the light sea quark mass. On the two $B$-fields we simulated for the $\Omega^-$, we see a slight $B^2$ dependence in the magnetic moment, roughly of the same size as for the $\Delta$. Additionally, we see that the errors associated with the experimental value are comparable to the statistical lattice errors here.

In order to improve on the quoted results, we must account for the systematic
errors that arise from a variety of sources in the calculation. First there is
the finite lattice spacing, which is difficult to estimate given the lack of
any calculations of the magnetic moments 
(quenched or dynamical) at multiple lattice spacings.  
Given we are using Clover fermions, errors of $O(a)$ disappear, so one would
expect errors to be roughly $O(a^2\Lambda_{\rm QCD}^2) \ltwid 0.03$. As it
will be some time before a second lattice spacing is available on 
these configurations, we will assume there is a 3\% systematic 
error that arises from the finite spatial lattice spacing. 

Additionally there are errors arising from remnant finite volume effects,
coming not from the background field, but from the pion mass. 
These are most likely negligible since in all cases, $m_\pi L \gtwid 4$, 
and thus the errors from these effects are less than $e^{-m_\pi L}\approx2\%$. 

Finally, there are uncertainties that plague any current calculation of the magnetic moments on the lattice, being that the sea quark charges are set to zero. We expect these to not be very large, coming from the discarded diagrams in which the valence quarks in the baryons emit photons that couple to the sea quarks. These at most are of order $\alpha$ relative to the terms that are included, and thus are expected to be at most $1\%$. Related errors are those coming from the $B^2$ extrapolation, and this is going to be at most 5\%, as mentioned below Eq.~(\ref{eq:B2extrap}).

We have presented here the first (using a background field method) dynamical
results for the $\Delta$ and $\Omega^-$ magnetic moments on dynamical
2+1-flavor lattices, which are consistent (given the pion mass used) with
experimental values that have been measured. Presently, the accuracy obtained
in the lattice result for the $\Omega^-$ magnetic dipole moment is comparable
with the experimental accuracy. We can use the above discussion to estimate the systematic error on our result for the $\Omega^-$ magnetic moment. We make a conservative estimate, and use the maximum values for each source of systematic error, and add those in quadrature, giving an error of 6\%. Thus we quote $\mu_{\Omega^-} = -1.93(8)(12)\mu_N$ for our final result.

To make significant progress on these results, simulations going to lighter pion masses, especially below the $\Delta\to \pi N$ threshold, are essential in precisely determining the magnetic moments. Nevertheless, it is rather encouraging that one can obtain already such precise results with the resources currently available.

We would like to thank NERSC and USQCD for the computing resources used to carry out this study, as well as the Jefferson Lab Lattice group for the anisotropic clover lattices. This work was partially supported by the US Department of Energy, under contract nos. DE-AC05-06OR23177 (JSA), DE-FG02-07ER41527, and DE-FG02-04ER41302; and by the Jeffress Memorial Trust, grant J-813. 

\bibliography{refs}

\begin{thebibliography}{13}
\expandafter\ifx\csname natexlab\endcsname\relax\def\natexlab#1{#1}\fi
\expandafter\ifx\csname bibnamefont\endcsname\relax
  \def\bibnamefont#1{#1}\fi
\expandafter\ifx\csname bibfnamefont\endcsname\relax
  \def\bibfnamefont#1{#1}\fi
\expandafter\ifx\csname citenamefont\endcsname\relax
  \def\citenamefont#1{#1}\fi
\expandafter\ifx\csname url\endcsname\relax
  \def\url#1{\texttt{#1}}\fi
\expandafter\ifx\csname urlprefix\endcsname\relax\def\urlprefix{URL }\fi
\providecommand{\bibinfo}[2]{#2}
\providecommand{\eprint}[2][]{\url{#2}}

\bibitem[{\citenamefont{Bernard et~al.}(1982)\citenamefont{Bernard, Draper,
  Olynyk, and Rushton}}]{Bernard:1982yu}
\bibinfo{author}{\bibfnamefont{C.~W.} \bibnamefont{Bernard}},
  \bibinfo{author}{\bibfnamefont{T.}~\bibnamefont{Draper}},
  \bibinfo{author}{\bibfnamefont{K.}~\bibnamefont{Olynyk}}, \bibnamefont{and}
  \bibinfo{author}{\bibfnamefont{M.}~\bibnamefont{Rushton}},
  \bibinfo{journal}{Phys. Rev. Lett.} \textbf{\bibinfo{volume}{49}},
  \bibinfo{pages}{1076} (\bibinfo{year}{1982}).

\bibitem[{\citenamefont{Cloet et~al.}(2003)\citenamefont{Cloet, Leinweber, and
  Thomas}}]{Cloet:2003jm}
\bibinfo{author}{\bibfnamefont{I.~C.} \bibnamefont{Cloet}},
  \bibinfo{author}{\bibfnamefont{D.~B.} \bibnamefont{Leinweber}},
  \bibnamefont{and} \bibinfo{author}{\bibfnamefont{A.~W.}
  \bibnamefont{Thomas}}, \bibinfo{journal}{Phys. Lett.}
  \textbf{\bibinfo{volume}{B563}}, \bibinfo{pages}{157} (\bibinfo{year}{2003}),
  \eprint{hep-lat/0302008}.

\bibitem[{\citenamefont{Lee et~al.}(2005)\citenamefont{Lee, Kelly, Zhou, and
  Wilcox}}]{Lee:2005ds}
\bibinfo{author}{\bibfnamefont{F.~X.} \bibnamefont{Lee}},
  \bibinfo{author}{\bibfnamefont{R.}~\bibnamefont{Kelly}},
  \bibinfo{author}{\bibfnamefont{L.}~\bibnamefont{Zhou}}, \bibnamefont{and}
  \bibinfo{author}{\bibfnamefont{W.}~\bibnamefont{Wilcox}},
  \bibinfo{journal}{Phys. Lett.} \textbf{\bibinfo{volume}{B627}},
  \bibinfo{pages}{71} (\bibinfo{year}{2005}), \eprint{hep-lat/0509067}.

\bibitem[{\citenamefont{Alexandrou et~al.}(2007)\citenamefont{Alexandrou,
  Korzec, Leontiou, Negele, and Tsapalis}}]{Alexandrou:2007we}
\bibinfo{author}{\bibfnamefont{C.}~\bibnamefont{Alexandrou}},
  \bibinfo{author}{\bibfnamefont{T.}~\bibnamefont{Korzec}},
  \bibinfo{author}{\bibfnamefont{T.}~\bibnamefont{Leontiou}},
  \bibinfo{author}{\bibfnamefont{J.~W.} \bibnamefont{Negele}},
  \bibnamefont{and} \bibinfo{author}{\bibfnamefont{A.}~\bibnamefont{Tsapalis}},
  \bibinfo{journal}{PoS} \textbf{\bibinfo{volume}{LAT2007}},
  \bibinfo{pages}{149} (\bibinfo{year}{2007}), \eprint{arXiv:0710.2744}.

\bibitem[{\citenamefont{Alexandrou et~al.}(2008)}]{Alexandrou:2008bn}
\bibinfo{author}{\bibfnamefont{C.}~\bibnamefont{Alexandrou}}
  \bibnamefont{et~al.} (\bibinfo{year}{2008}), \eprint{arXiv:0810.3976}.

\bibitem[{\citenamefont{Damgaard and Heller}(1988)}]{Damgaard:1988hh}
\bibinfo{author}{\bibfnamefont{P.~H.} \bibnamefont{Damgaard}} \bibnamefont{and}
  \bibinfo{author}{\bibfnamefont{U.~M.} \bibnamefont{Heller}},
  \bibinfo{journal}{Nucl. Phys.} \textbf{\bibinfo{volume}{B309}},
  \bibinfo{pages}{625} (\bibinfo{year}{1988}).

\bibitem[{\citenamefont{Rubinstein et~al.}(1995)\citenamefont{Rubinstein,
  Solomon, and Wittlich}}]{Rubinstein:1995hc}
\bibinfo{author}{\bibfnamefont{H.~R.} \bibnamefont{Rubinstein}},
  \bibinfo{author}{\bibfnamefont{S.}~\bibnamefont{Solomon}}, \bibnamefont{and}
  \bibinfo{author}{\bibfnamefont{T.}~\bibnamefont{Wittlich}},
  \bibinfo{journal}{Nucl. Phys.} \textbf{\bibinfo{volume}{B457}},
  \bibinfo{pages}{577} (\bibinfo{year}{1995}), \eprint{hep-lat/9501001}.

\bibitem[{\citenamefont{Aubin et~al.}(2008)\citenamefont{Aubin, Orginos,
  Pascalutsa, and Vanderhaeghen}}]{Aubin:2008hz}
\bibinfo{author}{\bibfnamefont{C.}~\bibnamefont{Aubin}},
  \bibinfo{author}{\bibfnamefont{K.}~\bibnamefont{Orginos}},
  \bibinfo{author}{\bibfnamefont{V.}~\bibnamefont{Pascalutsa}},
  \bibnamefont{and}
  \bibinfo{author}{\bibfnamefont{M.}~\bibnamefont{Vanderhaeghen}},
  \bibinfo{journal}{PoS} \textbf{\bibinfo{volume}{LAT2008}},
  \bibinfo{pages}{146} (\bibinfo{year}{2008}), \eprint{arXiv:0809.1629}.

\bibitem[{\citenamefont{Edwards et~al.}(2008)\citenamefont{Edwards, Joo, and
  Lin}}]{Edwards:2008ja}
\bibinfo{author}{\bibfnamefont{R.~G.} \bibnamefont{Edwards}},
  \bibinfo{author}{\bibfnamefont{B.}~\bibnamefont{Joo}}, \bibnamefont{and}
  \bibinfo{author}{\bibfnamefont{H.-W.} \bibnamefont{Lin}}
  (\bibinfo{year}{2008}), \eprint{arXiv:0803.3960}.

\bibitem[{\citenamefont{Lin et~al.}(2008)}]{Lin:2008pr}
\bibinfo{author}{\bibfnamefont{H.-W.} \bibnamefont{Lin}} \bibnamefont{et~al.}
  (\bibinfo{year}{2008}), \eprint{0810.3588}.

\bibitem[{\citenamefont{Stathopoulos and Orginos}(2007)}]{Stathopoulos:2007zi}
\bibinfo{author}{\bibfnamefont{A.}~\bibnamefont{Stathopoulos}}
  \bibnamefont{and} \bibinfo{author}{\bibfnamefont{K.}~\bibnamefont{Orginos}}
  (\bibinfo{year}{2007}), \eprint{arXiv:0707.0131}.

\bibitem[{\citenamefont{Pascalutsa and
  Vanderhaeghen}(2005)}]{Pascalutsa:2004je}
\bibinfo{author}{\bibfnamefont{V.}~\bibnamefont{Pascalutsa}} \bibnamefont{and}
  \bibinfo{author}{\bibfnamefont{M.}~\bibnamefont{Vanderhaeghen}},
  \bibinfo{journal}{Phys. Rev. Lett.} \textbf{\bibinfo{volume}{94}},
  \bibinfo{pages}{102003} (\bibinfo{year}{2005}), \eprint{nucl-th/0412113}.

\bibitem[{\citenamefont{Kotulla et~al.}(2002)}]{Kotulla:2002cg}
\bibinfo{author}{\bibfnamefont{M.}~\bibnamefont{Kotulla}} \bibnamefont{et~al.},
  \bibinfo{journal}{Phys. Rev. Lett.} \textbf{\bibinfo{volume}{89}},
  \bibinfo{pages}{272001} (\bibinfo{year}{2002}).

\end{thebibliography}

\end{document}